\documentclass[manuscript]{aastex}

\slugcomment{Not to appear in Nonlearned J., 45.}

\shorttitle{Estimating R-Process Yields from Abundances}
\shortauthors{Li et al.}

\begin{document}

%% LaTeX will automatically break titles if they run longer than
%% one line. However, you may use \\ to force a line break if
%% you desire.
\title{Estimating R-Process Yields from Abundances of the Metal-Poor Stars}

%% Use \author, \affil, and the \and command to format
%% author and affiliation information.
%% Note that \email has replaced the old \authoremail command
%% from AASTeX v4.0. You can use \email to mark an email address
%% anywhere in the paper, not just in the front matter.
%% As in the title, use \\ to force line breaks.

\author{Hongjie Li\altaffilmark{1,2}, Wenjuan Ma\altaffilmark{3}, Wenyuan Cui\altaffilmark{1} and Bo Zhang\altaffilmark{1,4}}

\affil{1. Department of Physics, Hebei Normal University, No. 20
East of South 2nd Ring Road, Shijiazhuang 050024, China \\
2. School of Sciences, Hebei University of Science and Technology,
Shijiazhuang 050018, China\\
3. Department of Physics, Cangzhou Normal University, Cangzhou
061001, China}

\altaffiltext{4}{Corresponding author. E-mail address:
zhangbo@mail.hebtu.edu.cn}

\begin{abstract}

The chemical abundances of metal-poor stars provide important clues
to explore stellar formation history and set significant constraints
on models of the r-process. In this work, we find that the abundance
patterns of the light and iron group elements of the main r-process
stars are very close to those of the weak r-process stars. Based on
a detailed abundance comparison, we find that the weak r-process
occurs in supernovae with a progenitor mass range of
$\sim11-26M_{\odot}$. Using the SN yields given by Heger \& Woosley
and the abundances of the weak r-process stars, the weak r-process
yields are derived. The SNe with a progenitor mass range of
$15M_{\odot}<M<26M_{\odot}$ are the main sites of the weak r-process
and their contributions are larger than 80\%. Using the abundance
ratios of the weak r-process and the main r-process in the solar
system, the average yields of the main r-process are estimated. The
observed correlations of the [neutron-capture/Eu] versus [Eu/Fe] can
be explained by mixing of the two r-process abundances in various
fractions.

\end{abstract}


\section{Introduction}

Heavy elements are created in slow (s-process) and rapid (r-process)
neutron-capture process \citep{bur57}. Although many authors thought
that the r-process sites are related to Type II supernovae (SNeII)
explosions \citep{cow06,arn07,sne08}, this has not yet been fully
confirmed. To investigate the r-process sites, the chemical
abundances of the metal-poor stars are important. Observed element
abundances of the ``main r-process stars" CS 22892-052 \citep{sne03}
and CS 31082-001 \citep{hil02} show that the heavy element
(Z$\geq56$) patterns are very similar to the scaled solar r-process
abundance pattern, while the lighter neutron-capture elements are
deficient in the solar r-process pattern \citep{cow06}. This implies
that the main r-process is not enough to explain the solar r-process
pattern. In contrast, observations of the very metal-poor stars HD
122563 ([Eu/Fe]$\approx-0.5$:\cite{wes00}) and HD 88609
\citep{hon07} show that there is an excess of their lighter
neutron-capture elements (e.g. Sr, Y and Zr). This indicates that
their abundances could come from another component: ``lighter
element primary process" \citep{tra04} or ``weak r-process
component" \citep{wan06,izu09}. \cite{mon07} have proposed that this
abundance pattern is uniform and unique.

Based on the abundance analysis, \cite{roe10a} found that the
abundances of other metal-poor stars seemed to lie in the continuum
between the patterns of the main r-process stars and the weak
r-process stars. In this case, they presented the idea that the
abundances of CS 22892-052 and HD 122563 could not be two standard
patterns of the r-process. They proposed that the two patterns may
represent the complete r-process and the incomplete r-process,
respectively, and suggested that the mixing of two patterns should
not be responsible for the large range of observed [Y/Eu] of the
metal-poor stars. Recently, \cite{boy12} proposed that the heavy
elements observed in some metal-poor stars, such as HD 122563, are
produced by the incomplete r-process, since the massive stars
collapse to black holes to truncate the r-process. They found that
the calculated result can not match the abundances of weak r-process
star HD 122563 and suggested that more exploration of truncated
r-process is needed.

The abundances of heavy elements and light elements in the extreme
metal-poor stars ([Fe/H]$\leq-2.5$) can provide significant clues
about r-process nucleosynthesis, because their abundances should
keep the abundance characteristics produced by a few SNe
\citep{mcw95a,mcw95b}. In this aspect, the main r-process stars
merit special attention. Their abundances could reflect results of
the main r-process nucleosynthesis that occurred in a SN. The very
high ratios of [neutron-capture/Fe] imply that the production of
main r-process elements does not couple with the iron group elements
\citep{qia07}. On the other hand, the abundances of weak r-process
stars should be close to the results for the weak r-process
nucleosynthesis. The ratios of [Sr/Fe]$\approx0$ mean that ejection
of weak r-process elements from a SN couple occurs with the ejection
of iron group elements. So, the abundances of weak r-process
elements should couple with the abundances of light elements and
iron group elements. Recently, \cite{li13c} derived the main
r-process and weak r-process components and used them to study the
stellar abundances \citep{li13a,li13b}.

Although the r-process sites can be studied by comparing the model
predictions with the observed abundances, the different studies
obtained different conclusions (e.g. \cite{tra99,ces08}). Recently,
\cite{mat14} studied the Eu yields in compact binary mergers (CBM)
and found that CBM should be responsible for Eu abundances in the
Galaxy. However, they reported that the time of the binary neutron
star mergers and the progenitor mass range of neutron stars are
still uncertain. Although many models have been presented, the
r-process nucleosynthesis sites producing the neutron-capture
elements of the metal-poor stars are still unknown. In this case,
the detailed analyses about the correlation between the abundances
of the light and iron group elements with those of the
neutron-capture elements in metal-poor stars should be important. In
this paper, we extract abundance clues by comparing the abundance
patterns between the main r-process stars and the weak r-process
stars in section 2. In Section 3, the progenitor mass ranges of the
weak r-process and main r-process are investigated. The quantitative
estimates for the average yields of the weak r-process and the main
r-process are presented in section 4. In section 5, the explanation
of correlations between [X$_{i}$/Eu] and [Eu/Fe] are given. Section
6 is our conclusions.

\section{Abundance Clues}

The $\alpha$ elements (e.g. Mg, Si, Ca and Ti) in the metal-poor
stars are definitely produced in massive stars \citep{woo95,heg10}.
To investigate the relationship between the abundances of $\alpha$
elements and main r-process elements, the observed abundance ratios
of [$\alpha$/Eu] as the function of [Eu/Fe] for metal-poor stars
\citep{wes00,cow02,hil02,joh02,sne03,chr04,hon04,bar05,iva06,lai08,hay09,mas10,roe10b}
are shown in Figure 1. Clearly, the relationships of [$\alpha$/Eu]
versus [Eu/Fe] in Figure 1 are close to straight lines with slopes
of approximately -1. This indicates that the $\alpha$ elements are
not correlated with Eu. This noncorrelations imply that the mass
ranges of the massive stars producing the $\alpha$ elements are
different from those of the progenitors of SNe II from which the
main r-process elements are ejected. Furthermore, the observations
also mean that the gas clouds in which main r-process stars formed
have been polluted by the nucleosynthesis process producing the
$\alpha$ elements in the massive stars.

Although the r-process sites have not been fully confirmed, much
evidence suggests that the r-process is related to SNe II from
massive stars \citep{sne08}. For investigating the mass range of the
progenitor in which the r-process occurs, the stellar abundances
mainly polluted by one process, such as the abundances of the light
elements and iron group elements in main r-process stars and weak
r-process stars, are significant since they can be compared with the
nucleosynthesis calculations.

In Figure 2, the comparisons of the abundances of the weak r-process
stars \citep{hon04,hon06,hon07} with those of the main r-process
stars \citep{hil02,sne03,sne08} are shown, in which the abundances
of HD 88609, CS 22892-052 and CS 31082-001 have been normalized to
the Fe abundance of HD 122563. Obviously, the weak r-process stars
HD 122563 and HD 88609 have similar abundance patterns: excesses of
light neutron-capture elements and underabundances of heavy
neutron-capture elements, which is different than the abundance
patterns of the main r-process stars. Although the abundance
patterns of neutron-capture elements for the weak r-process stars
and main r-process stars are obviously different, it is noteworthy
that their abundance patterns for light elements and iron group
elements are close to each other. In this case, we first normalize
the abundances of the light elements and iron group elements of HD
88609, CS 22892-052 and CS 31082-001 to those of HD 122563 and
derive the average abundance pattern of four sample stars. Then, the
abundances of the light elements and iron group elements of the four
stars are normalized to the average abundance pattern. The top panel
of Figure 3 shows the average abundance pattern and the best-fit
results for four sample stars. There is good agreement between the
average abundance pattern and the observed data for the four stars
from O to Zn. The rms offsets of these elements are shown in the
middle panel. Obviously, the average abundances are a good,
representative pattern of the light elements and iron group elements
for the weak r-process stars and main r-process stars. Because the
abundance patterns of these stars are mainly polluted by a few
nucleosynthesis events, the similarity in the abundance patterns of
light elements and iron group elements between the weak r-process
stars and main r-process stars means that the abundance pattern is
stable and universal. Although the abundances of neutron-capture
elements are different obviously, the astrophysical origins of the
light elements and iron group elements in these two kinds of stars
should be similar.

\section{The Range of Progenitor Mass}

The weak r-process and main r-process are associated with a
core-collapse SNe explosion, but the astrophysical sites have not
yet been fully confirmed. Whether or not the light and iron group
elements are ejected by a core-collapse SNe relate to the
progenitors' masses. Two mass ranges of the massive stars should
lead to a core-collapse SNe: the O-Ne-Mg core-collapse SNe with an
initial mass of $8-10M_{\odot}$ and the Fe core-collapse SNe with an
initial mass of $11-25M_{\odot}$ \citep{qia07}. Our goal is to
investigate the mass range of the progenitors of the SNe in which
the weak r-process occurred. To find the mass range of the
progenitors producing the abundance pattern of light and iron group
elements of the weak r-process stars, we use the single SN yields
presented by \cite{heg10} to fit the average abundances of four
stars (from O to Zn). The best-fit result is shown in the bottom
panel of Figure 3. We find that the average abundances are best
matched by a progenitor mass of $23 M_{\odot}$ with
$\chi^{2}=0.864$. Taking a mass interval of $0.5 M_{\odot}$ for
$M<30M_{\odot}$, Figure 4 displays the calculated lower limit of
$\chi^{2}$ as a function of the progenitor mass. Obviously, the
lower limit of $\chi^{2}$ is sensitive to the progenitor mass. The
average abundance pattern can be fitted by the yields of the massive
stars of $11M_{\odot}<M<26M_{\odot}$, with $\chi^{2}<2$. The fitted
results are significant evidence that the weak r-process occurs in
the supernova with the progenitor masses of $\sim11-26M_{\odot}$.
Furthermore, because the sites producing the main r-process elements
do not produce light and iron group elements, this process must
occur in the supernova with progenitor masses of
$\sim8-10M_{\odot}$. \cite{boy12} have suggested that the r-process
elements may be produced in the massive stars with $8-40M_{\odot}$.
The fitted results lie in their progenitor mass range.

\section{Estimating the R-process Yields}

Using the light element yields $Y_{l}$ of a single SN calculated by
\cite{heg10} and the weak r-process abundance pattern presented by
\cite{li13c}, it was possible to derive the weak r-process yields.
The relationship between weak r-process yields $Y_{r,w}$ and weak
r-process abundance $N_{r,w}$ is
\begin{equation}
Y_{r,w}/Y_{l}=N_{r,w}\times{A_{r,w}}/(N_{l}\times{A_{l}})
\end{equation}
where $A$ is atomic weight. Combining the light element yields
$Y_{l}$ calculated by \cite{heg10} and equation (1), we obtain the
Sr yields as a function of progenitor mass, which are shown in the
top panel in Figure 5. The Sr yields increase with the progenitor
mass and reach a maximum at about $1.33\times10^{-5}M_{\odot}$ at
$M=24.5M_{\odot}$. Assuming a Salpeter initial mass function (IMF),
the normalized fractions of Sr yields in weak r-process per unit
mass interval (in units of solar mass) compared to the total
contribution from $10.5-26.5M_{\odot}$ are shown in the bottom panel
of Figure 5. The calculated results indicate that the SNe with a
progenitor mass range of 15$M_{\odot}<M<26M_{\odot}$ are the main
sites of the weak r-process. The contributions of the SNe to the
abundances of the weak r-process elements in our galaxy are larger
than 80\%. The average weak r-process yields produced by one weak
r-process event, which have been weighted by the Salpeter IMF, are
listed in Table 1.

Because the solar r-process abundances can be matched by main
r-process and weak r-process abundances obtained from the metal-poor
stars \citep{li13c}, the two r-process patterns are independent of
metallicity. In this case, the solar r-process pattern has been
divided into two components. The r-weak:r-main ratios for Sr, Y and
Zr are about 0.66:0.34, 0.74:0.26 and 0.73:0.27, respectively.
Assuming the Salpeter IMF, the number ratio of the massive stars
with $10.5-26.5M_{\odot}$ to the massive stars with $8-10M_{\odot}$
is about 1.9. Taking Sr as a representative element and using the
average Sr yield listed in Table 1, the average Sr yield of one main
r-process event is derived to be about $3.06
\times10^{-6}M_{\odot}$. Adopting this method, we obtained the
average yields of the other neutron-capture elements, which are
listed in Table 2. In Figures 6 and 7, the estimated yields of the
weak r-process versus the progenitor mass are plotted by curves. The
straight lines represent the average yields of the main r-process.
For convenient comparison, the dash lines divide each picture into
two parts. Our calculated results are based on the assumption that
the main r-process occurs in the SN II. Note that \cite{mat14} have
suggested that CBM should be responsible for Eu abundances in the
Galaxy. Considering the contributions of the CBM to the abundances
of the r-process elements, the yields of the r-process listed in the
Table 1 and Table 2 should be the upper limits of the yields
produced in the SNe II. Obviously, more investigations on this
subject are needed.

\section{Explanations of the Correlations between [X$_{i}$/Eu] and [Eu/Fe]}

\cite{mon07} have found that the slopes of [X$_{i}$/Eu] versus
[Eu/Fe] for the lighter neutron-capture elements are about -1 for
[Eu/Fe]$\lesssim$1.0. However, the ratios of the [X$_{i}$/Eu]
flattened for higher [Eu/Fe]. Recently, \cite{boy12} suggested that
the heavy elements of some metal-poor stars, such as HD 122563, are
produced by the ``truncated r-process", because the more massive
stars collapse to black holes before the r-process is completed.
However, their calculated result demonstrates that the truncated
r-process predictions can only explain the downward abundance trend
as atomic number increases and cannot match the abundances of HD
122563. Obviously, more exploration of the truncated r-process is
needed \citep{boy12}.

The average ratios [Sr/Fe] of weak r-process stars and main
r-process stars are about -0.15 and 0.6 respectively. However, there
are some low-Sr stars ([Sr/Fe]$\lesssim$-1) in the metal-poor stars
(see Fig. 8 in \cite{han12}), which means that there is another
component barely producing neutron-capture elements in the early
Galaxy. This component was called the prompt (P) component
\citep{qia01}. Once the SNe II in which r-process elements are
produced began to pollute the interstellar medium, the effect of the
P component became smaller \citep{qia01}. In this section, we wish
to explain the observed trends of [X$_{i}$/Eu] versus [Eu/Fe]
quantitatively using derived yields of the two r-processes, so our
sample stars do not contain the low-Sr stars.

Figure 8 shows the relationships between [X$_{i}$/Eu] and [Eu/Fe],
where X$_{i}$ are the abundances of lighter neutron-capture
elements. The filled squares are the observed ratios of the
metal-poor stars
\citep{wes00,cow02,hil02,joh02,sne03,chr04,hon04,bar05,hon06,iva06,fra07,hon07,lai08,hay09,mas10,roe10a,roe10b}.
The dash lines correspond to the average abundance ratios of weak
r-process stars, whose slopes are -1. The dotted lines represent the
abundance ratios polluted by pure main r-process material. For
comparison, we added the dash dotted lines to represent the solar
r-process ratios, which are adopted from \cite{arl99} and updated
from \cite{tra04} for the ratios of Sr-Nb. Obviously, the large
scatter of abundance ratios [X$_{i}$/Eu] of the metal-poor stars
cannot be explained by the corresponding solar r-process ratios. We
find that the observed abundance ratios are close to the dash lines
for [Eu/Fe]$\lesssim$0. The reason of the abundance ratios decrease
linearly with increasing [Eu/Fe] for [Eu/Fe]$\lesssim$0 is that
these elements mainly come from the weak r-process. Obviously, the
observed abundance ratios are close to the main r-process lines but
not the weak r-process lines for [Eu/Fe]$>$1.0. The reason for
flattened [X$_{i}$/Eu] at higher [Eu/Fe] is that more contributions
come from the main r-process. In order to explain the observed
trends of the abundance ratios, we calculated the mixing of the weak
r-process abundances and the main r-process abundances with
different proportions, which are plotted by the solid lines. We can
see that the mixing lines are perfectly consistent with the
abundance trends. The results mean that the abundance trends can be
explained by the contributions of two r-processes. The fractions of
the main r-process that contributed to the abundances of the lighter
neutron-capture elements for the various [Eu/Fe] plotted by open
circles in Figure 8 are listed in Table 3. We can see that, for the
weak r-process stars, the fractions of the weak r-process that
contributed to lighter neutron-capture elements (from Sr to Ag) lie
in the range of 87\%-97\%. On the other hand, for the main r-process
stars, the contributed fractions of the main r-process are larger
than 77\%. Based on calculations of the r-process triggered by SNe
II explosions, \cite{han12} showed in their Table 3 that the
percentages contributed by the main r-process increase with the
increasing atomic number for a given electron fraction. From Table
3, we can see that the fractions contributed by the main r-process
to the abundances of the neutron-capture elements increase as the
atomic number increases for a given [Eu/Fe]. The results should
suggest that the increasing trends are the common phenomenon for the
low metallicity.

\section{Conclusions}

The abundances, especially the abundance patterns, of metal-poor
stars can provide important constraints on the r-process sites. In
this aspect, the detailed abundance analysis approach might provide
some helpful clues to how the core-collapse SNe relate to the
r-process. Our results can be summarized:

1. The slopes of [$\alpha$/Eu] versus [Eu/Fe] are roughly consistent
with -1. This indicates that the abundances of $\alpha$ elements do
not related to those of the main r-process elements. This
noncorrelation implies that the mass range of the massive stars in
which the $\alpha$ elements are produced is different from the mass
range of the progenitor of the SNe II in which the main r-process
occurs.

2. The abundance patterns of light and iron group elements of the
main r-process stars are very close to those of weak r-process
stars. This indicates that, although the ratios of [Eu/Fe] are
obviously different, the light and iron group elements of main
r-process stars and those of weak r-process stars should come from
massive stars with similar mass range. So, the abundances of the
main r-process stars also contain contributions from the weak
r-process. The difference in Eu abundances between the weak
r-process stars and the main r-process stars is mainly due to a
different polluted level of the main r-process.

3. The calculated results imply that the weak r-process occurs in
the supernovae with a progenitor mass range of
$\sim$11-26$M_{\odot}$. The SNe with progenitor mass range of
$15M_{\odot}<M<26M_{\odot}$ are the main origins of the weak
r-process elements. The average yields of one weak r-process event
had been derived.

4. The main r-process elements are produced in the supernovae with
progenitor masses of $\sim8-10M_{\odot}$. Using the contributed
ratios of the weak r-process and the main r-process to the solar
system, the average yields of one main r-process event are
estimated.

5. For the weak r-process stars, the fractions of the weak r-process
that contributed to lighter neutron-capture elements lie in the
range of 87\%-97\%. For the main r-process stars, the contributed
fractions of the main r-process are larger than 77\%. The observed
correlations between the [neutron-capture/Eu] versus [Eu/Fe] can be
explained by the mixing of the weak r-process abundances and the
main r-process abundances.

Our results could present some constraints for detailed r-process
models. Obviously, more detailed studies about the weak r-process
and the main r-process are needed.

\acknowledgments

We are most grateful to the referee for the insightful and
constructive suggestions, which improved this paper greatly. This
work has been supported by the National Natural Science Foundation
of China under Grant Nos. 11273011, U1231119, 10973006 and 11003002,
the Science Foundation of Hebei Normal University under Grant No.
L2009Z04, the Natural Science Foundation of Hebei Province under
Grant Nos. A2009000251, A2011205102, the Science and Technology
Supporting Project of Hebei Province under Grant No. 12211013D and
the Program for Excellent Innovative Talents in University of Hebei
Province under Grant No. CPRC034.

\clearpage

\begin{figure}[t]
 \centering
 \includegraphics[width=1\textwidth,height=0.6\textheight]{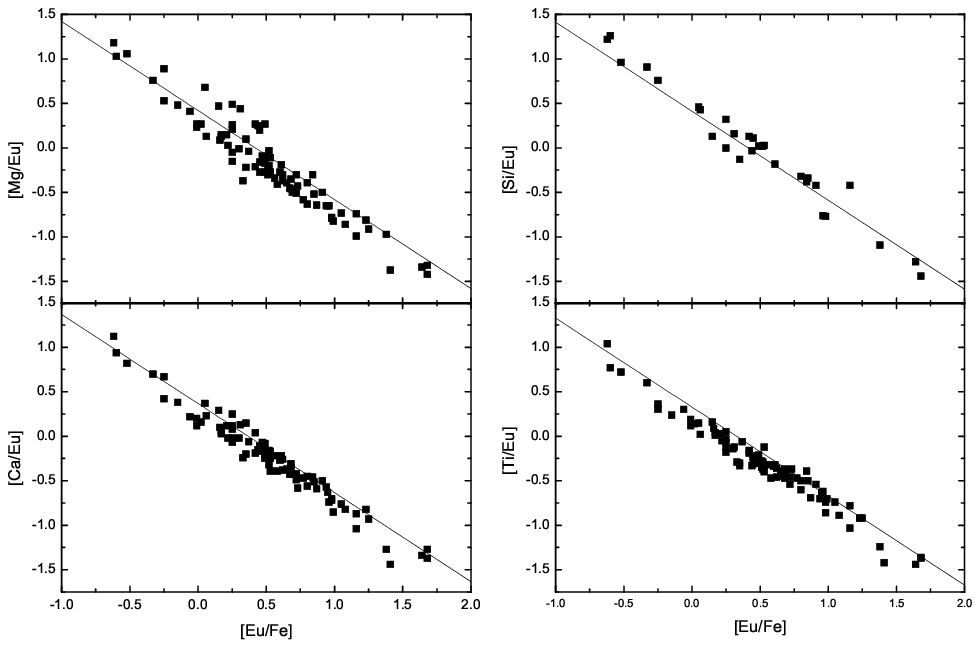}
 %\suppressfloats[t]
\caption{Plots of [$\alpha$/Eu] versus [Eu/Fe]. The filled squares
represent the observed abundances of the $\alpha$ elements from
\cite{wes00,cow02,hil02,joh02,sne03,chr04,hon04,bar05,iva06,lai08,hay09,mas10,roe10b}.
The slopes of the solid lines are -1.}
 %\label{appenfig}
\end{figure}

\begin{figure}[t]
 \centering
 \includegraphics[width=1\textwidth,height=0.6\textheight]{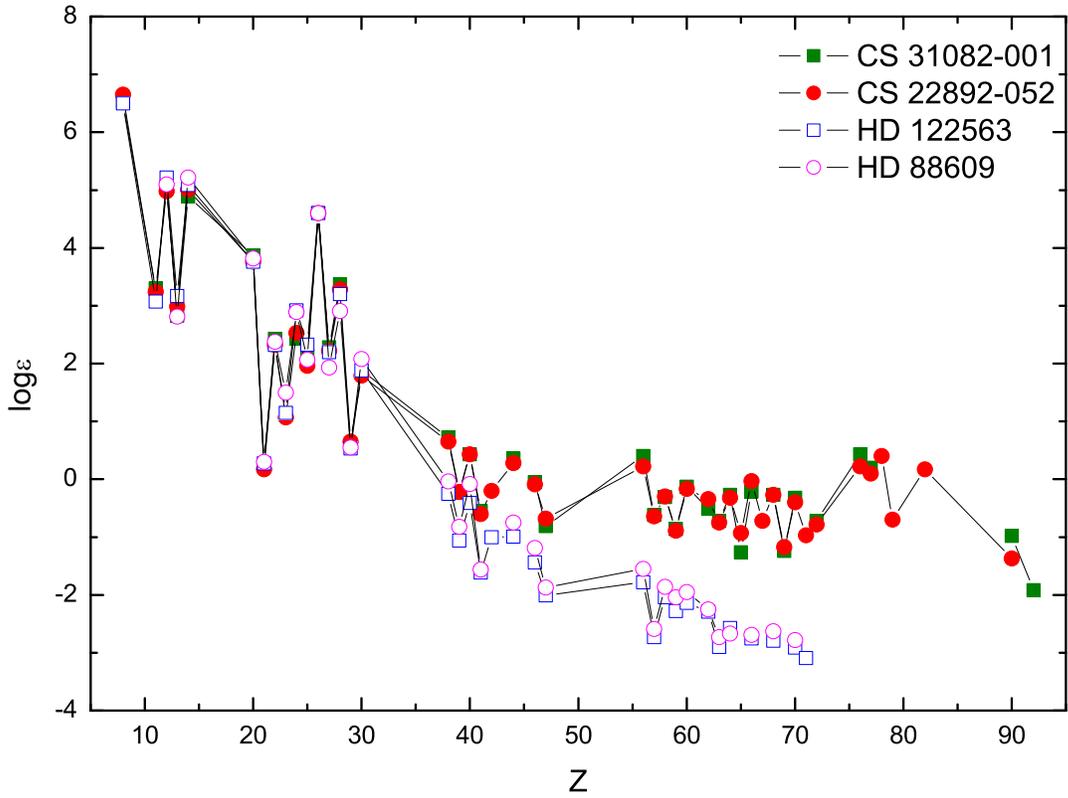}
 %\suppressfloats[t]
\caption{Comparisons of the abundance patterns of the weak r-process
stars HD 122563 (open squares) and HD 88609 (open circles) with
those of the main r-process stars CS 22892-052 (filled circles) and
CS 31082-001 (filled squares) on a logarithmic scale. The abundances
of the other three stars have been normalized to the Fe abundance of
HD 122563.}
 %\label{appenfig}
\end{figure}

\begin{figure}[t]
 \centering
 \includegraphics[width=1\textwidth,height=0.6\textheight]{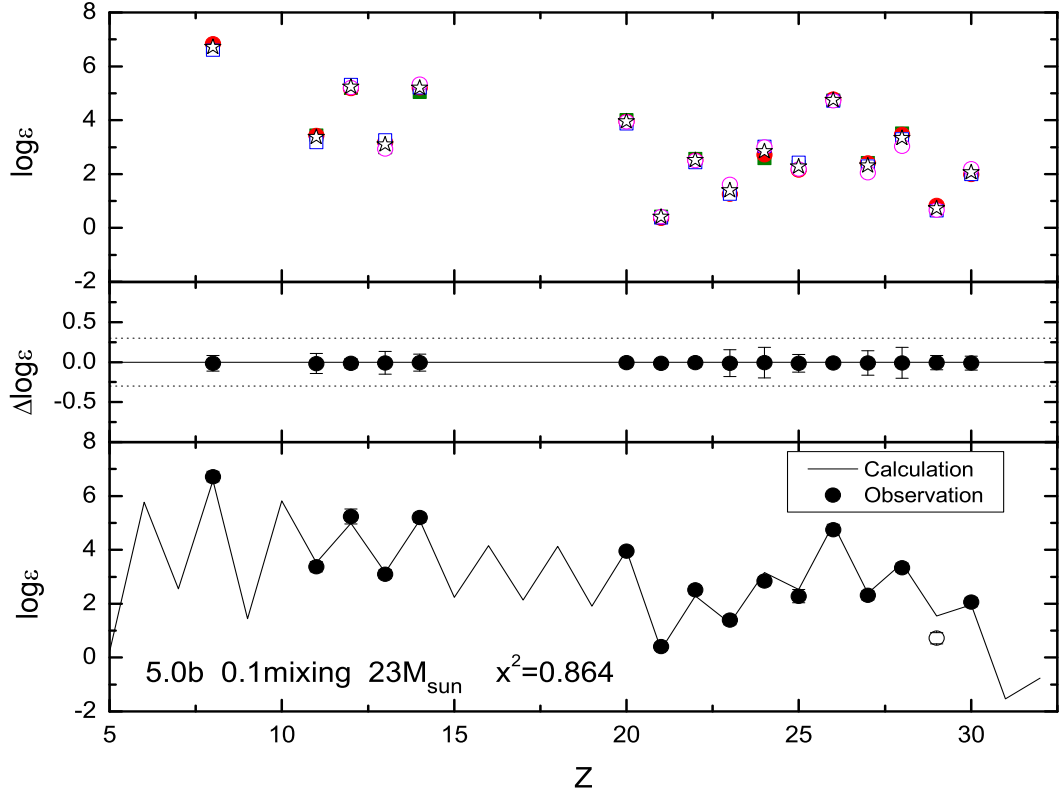}
 %\suppressfloats[t]
\caption{Top panel: Average abundance pattern and best-fit results
of four sample stars. Middle panel: the rms offset of these elements
in $log\varepsilon$. Bottom panel: Fitted average abundance pattern.
The symbol: the symbols for the four sample stars are the same as in
Fig. 2; the open stares are the average abundances. Typical
observational uncertainties in $log\varepsilon$ are $\sim0.2-0.3$
dex (dotted lines).}
 %\label{appenfig}
\end{figure}

\begin{figure}[t]
 \centering
 \includegraphics[width=1\textwidth,height=0.6\textheight]{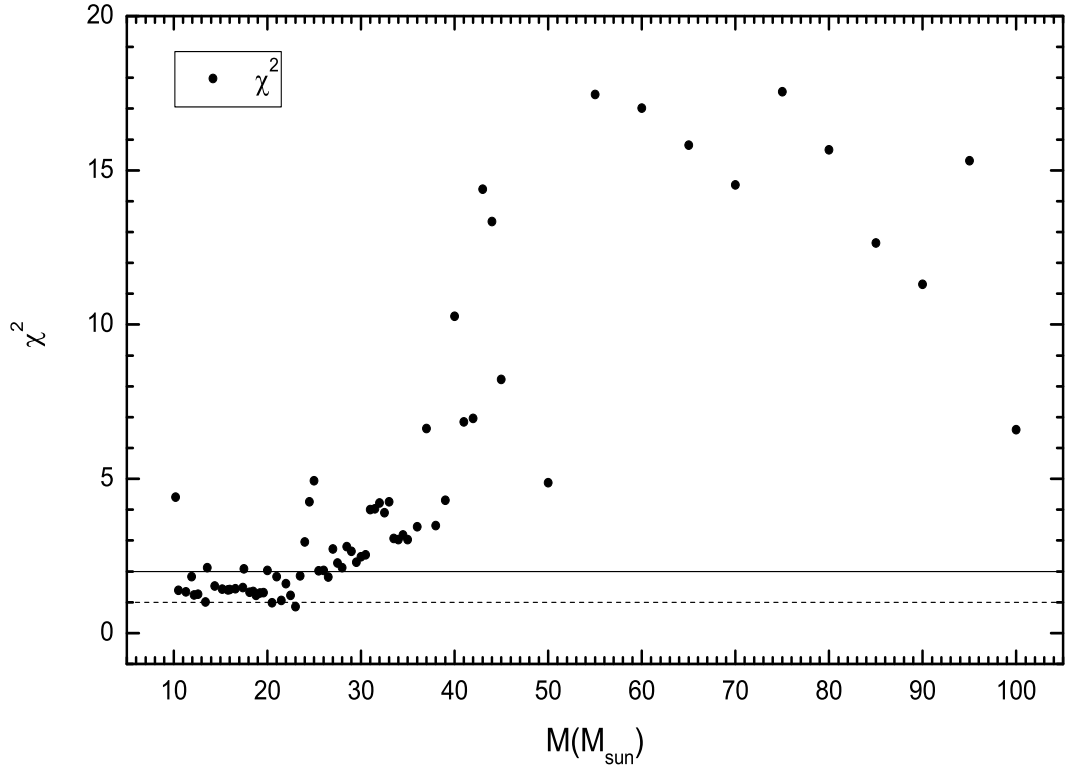}
 %\suppressfloats[t]
\caption{Calculated lower limit of $\chi^{2}$ as a function of the
progenitor mass. The dash line and solid line represent $\chi^{2}=1,
2$, respectively.}
 %\label{appenfig}
\end{figure}

\begin{figure}[t]
 \centering
 \includegraphics[width=1\textwidth,height=0.6\textheight]{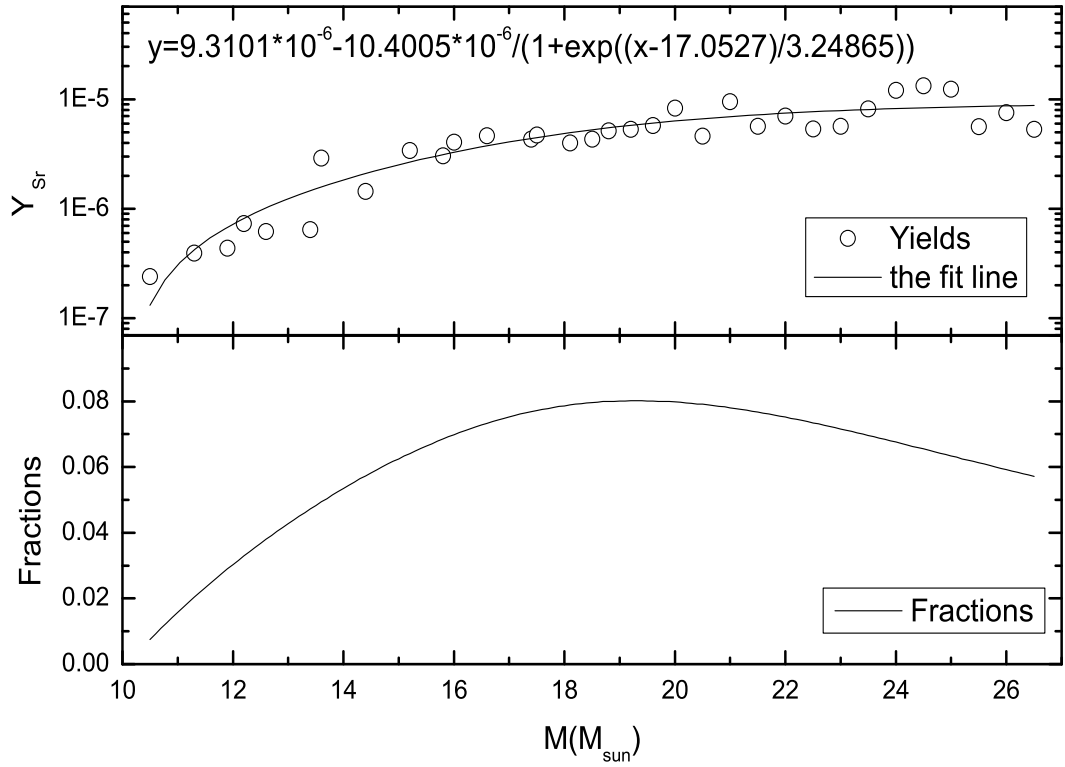}
 %\suppressfloats[t]
\caption{Top panel: Sr yields as a function of the progenitor mass
(solid line is the fit line). Bottom panel: Normalized fractions of
Sr yields in the weak r-process per unit interval of the progenitor
mass, compared to the total contribution from $10.5-26.5M_{\odot}$.}
 %\label{appenfig}
\end{figure}

\begin{figure}[t]
 \centering
 \includegraphics[width=1\textwidth,height=0.6\textheight]{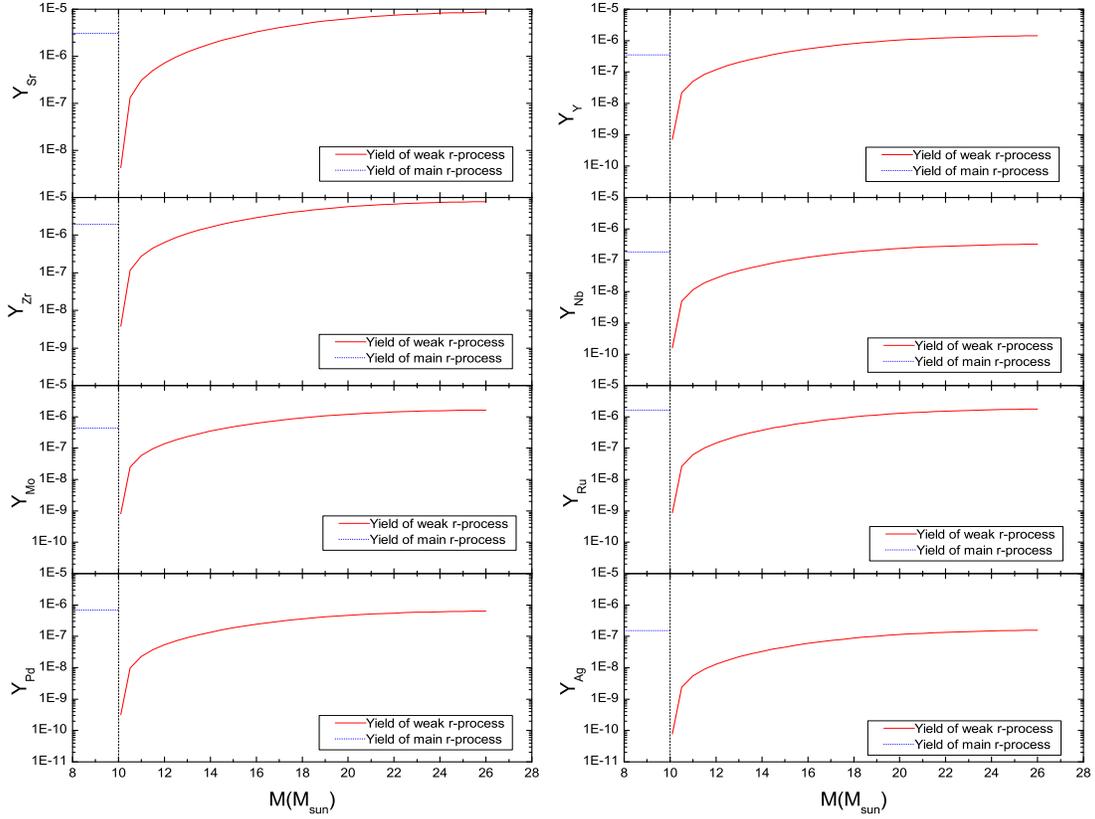}
 %\suppressfloats[t]
\caption{Estimate yields of the weak r-process and average yields of
the main r-process for lighter neutron-capture elements vs. the
progenitor mass. The straight lines (dotted lines) represent the
average yields of the main r-process. The curves (solid lines) are
the estimated yields of the weak r-process. For convenient
comparison, the dash lines divide each picture into two parts.}
 %\label{appenfig}
\end{figure}

\begin{figure}[t]
 \centering
 \includegraphics[width=1\textwidth,height=0.6\textheight]{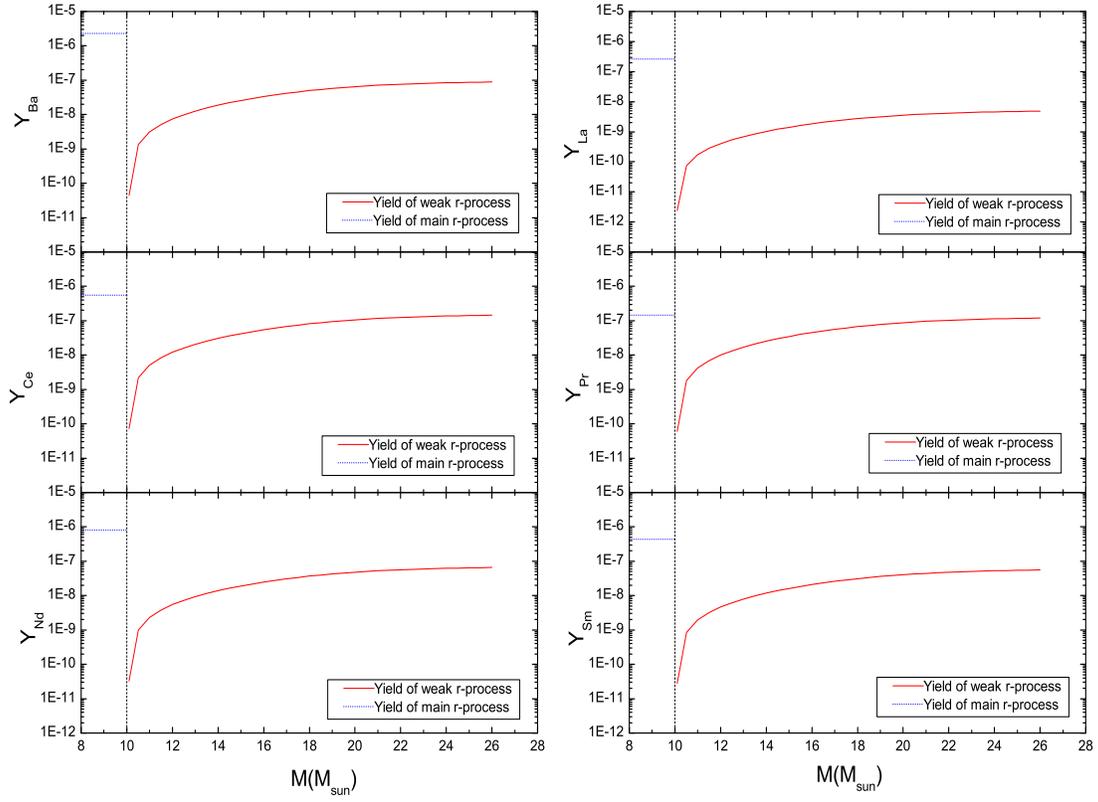}
 %\suppressfloats[t]
\caption{Estimated yields of the weak r-process and average yields
of the main r-process for heavy neutron-capture elements vs. the
progenitor mass. The symbols are the same as in Figure 6.}
 %\label{appenfig}
\end{figure}

\begin{figure}[t]
 \centering
 \includegraphics[width=1\textwidth,height=0.6\textheight]{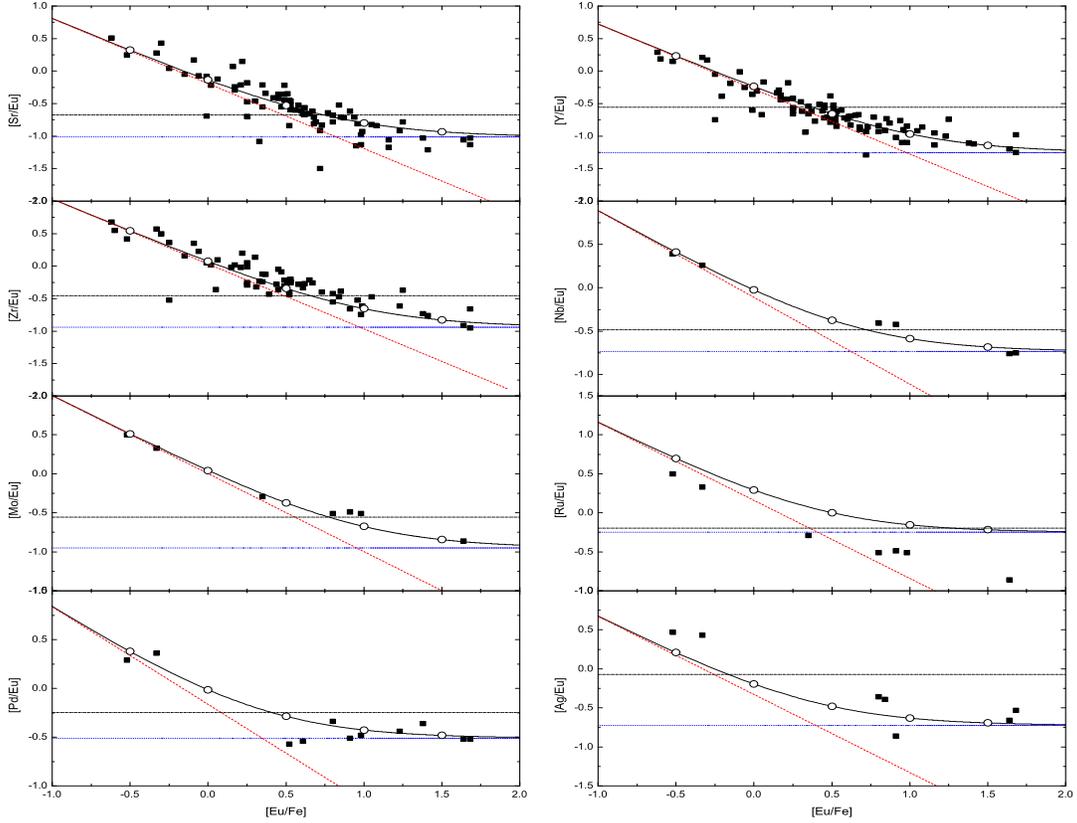}
 %\suppressfloats[t]
\caption{Plots of [X$_{i}$/Eu] vs. [Eu/Fe]. The filled squares
represent the observed abundances of these elements. The slopes of
the dashed lines are -1. The dotted lines represent the abundance
ratios enriched by pure main r-process material. The dash-dotted
lines represent the r-process ratios of the solar system. The curves
are a mixture of contributions from the weak r-process and the main
r-process. The open circles correspond to various [Eu/Fe].}
 %\label{appenfig}
\end{figure}

\clearpage

\begin{table}
\begin{center}
\caption{Average yields produced by the weak r-process
event.\label{tbl-1}}

\begin{tabular*}{458pt}{@{\extracolsep\fill}cccccc}

\tableline\tableline Z  &   element &   Yield($M_{\odot}$)  &     Z   &    element    &   Yield($M_{\odot}$)  \\
\tableline
38  &    Sr     &   3.14E-06 &   47  &    Ag     &   5.68E-08 \\
39  &    Y  &   5.18E-07 &   56  &    Ba     &   3.21E-08 \\
40  &    Zr     &   2.81E-06 &   57  &    La     &   1.75E-09 \\
41  &    Nb     &   1.18E-07  &   58  &    Ce     &   5.25E-08 \\
42  &   Mo  &   5.94E-07 &   59  &    Pr     &   4.34E-08 \\
44  &    Ru     &   6.35E-07 &   60  &    Nd     &   2.38E-08 \\
46  &    Pd     &   2.33E-07  &   62  &    Sm     &   2.02E-08 \\

\tableline
\end{tabular*}
\clearpage
\end{center}
\end{table}

\begin{table}
\begin{center}
\caption{Average yields produced by the main r-process
event.\label{tbl-2}}

\begin{tabular*}{458pt}{@{\extracolsep\fill}cccccc}

\tableline\tableline Z  &   element &   Yield($M_{\odot}$)  &     Z   &    element    &   Yield($M_{\odot}$)  \\
\tableline
38  &    Sr     &   3.06E-06    &   65  &   Tb  &   1.13E-07    \\
39  &    Y  &   3.52E-07    &   66  &    Dy     &   1.03E-06    \\
40  &    Zr     &   1.92E-06    &   67  &   Ho  &   2.64E-07    \\
41  &    Nb     &   1.83E-07    &   68  &    Er     &   7.33E-07    \\
42  &   Mo  &   4.30E-07    &   69  &    Tm     &   8.61E-08    \\
44  &    Ru     &   1.64E-06    &   70  &   Yb  &   5.79E-07    \\
46  &    Pd     &   6.97E-07    &   71  &   Lu  &   1.58E-07    \\
47  &    Ag     &   1.52E-07    &   72  &   Hf  &   2.89E-07    \\
56  &    Ba     &   2.30E-06    &   76  &   Os  &   3.46E-06    \\
57  &    La     &   2.64E-07    &   77  &    Ir     &   2.47E-06    \\
58  &    Ce     &   5.56E-07    &   78  &   Pt  &   4.14E-06    \\
59  &    Pr     &   1.46E-07    &   79  &   Au  &   3.30E-07    \\
60  &    Nd     &   8.24E-07    &   82  &   Pb  &   2.58E-06    \\
62  &    Sm     &   4.29E-07    &   90  &    Th     &   1.37E-07    \\
63  &    Eu     &   2.27E-07    &   92  &   U   &   2.24E-08    \\
64  &   Gd  &   7.28E-07    &       &       &       \\
\tableline
\end{tabular*}
\clearpage
\end{center}
\end{table}

\begin{table}
\begin{center}
\caption{Contributed fractions of the main r-process to the
abundances of lighter neutron-capture elements for various
[Eu/Fe].\label{tbl-3}}

\begin{tabular*}{458pt}{@{\extracolsep\fill}ccccccccc}

\tableline\tableline [Eu/Fe]  &   Sr &   Y  &     Zr   &    Nb    &   Mo  &  Ru  &  Pd  &  Ag  \\
\tableline
-0.5    &   0.05    &   0.03    &   0.03    &   0.07    &   0.03    &   0.11    &   0.13    &   0.12    \\
0   &   0.13    &   0.10    &   0.10    &   0.19    &   0.10    &   0.29    &   0.32    &   0.29    \\
0.5 &   0.32    &   0.25    &   0.25    &   0.43    &   0.26    &   0.56    &   0.60    &   0.57    \\
1   &   0.60    &   0.51    &   0.52    &   0.71    &   0.53    &   0.80    &   0.82    &   0.81    \\
1.5 &   0.83    &   0.77    &   0.77    &   0.88    &   0.78    &   0.93    &   0.94    &   0.93    \\
\tableline
\end{tabular*}
\clearpage
\end{center}
\end{table}

\end{document}